\documentclass[12pt, preprint]{aastex}

\renewcommand\th{\thinspace}
\newcommand\kms{\ifmmode{\rm km\th s^{-1}}\else km\th s$^{-1}$\fi}

\newcommand{\h}{$^h$~}
\newcommand{\m}{$^m$~}
\newcommand{\s}{$^s$}
\newcommand{\dg}{$^\circ$~}

\newcommand{\ha}{H$\alpha$}

\newcommand{\bti}{0.66^{+0.03}_{-0.09}}
\newcommand{\bri}{0.05^{+0.01}_{-0.02}}
\newcommand{\uplimone}{0.019}
\newcommand{\uplimtwo}{0.031}

\newcommand{\phihi}{1.6}
\newcommand{\haiso}{0.03}

\shorttitle{Extended \ha~in M31}
\shortauthors{Madsen et al. 2001}

\begin{document}

\title{Observations of the Extended Distribution of Ionized Hydrogen
  in the Plane of M31}
\author{G. J. Madsen, R. J. Reynolds, L. M. Haffner}
\affil{Department of Astronomy, University of Wisconsin--Madison, 475 
  North Charter Street, Madison, WI 53706-1582}
\email{madsen@astro.wisc.edu; reynolds@astro.wisc.edu; haffner@astro.wisc.edu}
\author{S. L. Tufte}
\affil{Department of Physics, Lewis \& Clark College, 0615 SW Palatine
  Hill Road, Portland, OR 97219}
\email{tufte@lclark.edu}
\author{P. R. Maloney}
\affil{CASA, Campus Box 389, University of Colorado, Boulder, CO
  80309}
\email{maloney@casa.colorado.edu}

\begin{abstract}
We have used the Wisconsin H-Alpha Mapper (WHAM) to observe the
spatially extended distribution of ionized hydrogen in M31 beyond the
stellar disk. We obtained five sets of observations, centered near
the photometric major axis of M31, that extend from the center of the
galaxy to just off the edge of the southwestern HI disk. Beyond the
bright stellar disk, but within the HI disk, weak \ha~is detected with
an intensity $I_{H\alpha} = \bri R$\footnotemark
\footnotetext{$1R=\frac{10^6}{4\pi}~$photons~cm$^{-2}$~s$^{-1}$~ster$^{-1}$}.
Since M31 is inclined
77\dg  with respect to the line of sight, this implies that the
ambient intergalactic ionizing flux onto each side of M31 is $\Phi_0 \leq 
\phihi \times 10^{4}~$~photons~cm$^{-2}$~s$^{-1}$. Just beyond the
outer boundary of the HI disk we find no significant detection of
\ha~and place an upper limit $I_{H\alpha} \leq \uplimone R$.
\end{abstract}

\keywords{ISM:general---intergalactic medium---galaxies:
  ISM---galaxies: individual (M31)}

\section{Introduction}

The distribution of neutral hydrogen (HI) in spiral galaxies often extends 
well beyond their optical disks (see review by \citet{san87}). 
Additionally, some spiral systems have neutral hydrogen column
densities that sharply truncate at large galactocentric
distances \citep{van91,cor93}.
This may be interpreted as the edge of the gaseous disk of the galaxy;
however, recent theoretical arguments and observational 
evidence suggest that an additional phase of material, ionized
hydrogen, may exist beyond the regions of observed HI \citep{mal93,blan97}.
Direct observations of this ionized gas would yield important physical
information about the true size of spiral galaxies, and with
sufficient sensitivity and spectral resolution, 
could extend the rotation curve of spiral galaxies, yielding more
insight into the dynamics of spiral galaxies and their distribution of
dark matter.
Observations of ionized hydrogen beyond the stellar disk may also lead
to information about the sources of ionization, which may be within
the galactic system, as in the case of a hot galactic corona, or
outside of it.
An external source of radiation, the intergalactic Lyman continuum
flux, plays an important role in cosmological models.
By observing the brightness of the \ha~in the outer HI disk of
a local galaxy, one may constrain the magnitude of the 
$z=0$ ionizing background flux.

We have used the
Wisconsin H-Alpha Mapper (WHAM) to observe the distribution
of diffuse ionized hydrogen in the local spiral galaxy M31. The high
sensitivity and large field of view of WHAM make M31 the best galaxy
in which to search for an extended disk of ionized hydrogen.

\section{Observations}

The WHAM instrument is a fully remotely operated facility with a
15 cm, dual-etalon Fabry-Perot spectrometer at the focal plane of a 0.6 m
telescope atop Kitt Peak in Arizona \citep{rey98,tuf97}.
The WHAM spectrometer has a 1$^\circ$
diameter circular field of view on the sky, and a velocity
resolution of 12 \kms~within a 200 \kms~wide spectral window that can
be centered on any wavelength between 4800 \AA~and 7300 \AA. WHAM was
designed to detect very weak emission lines from ionized gas.

For the observations of M31, the spectral window was centered at
approximately -500 \kms~with respect to rest \ha, which is the
approximate radial velocity of the southwestern half of the galaxy.
Spectra were obtained toward a total of nine lines of sight.
Five lines of sight (``ONs'') were centered near the photometric major
axis of M31 and extended from the center of M31 to just off the edge
of the observed southwestern HI disk (see Figure 1).
Four lines of sight (``OFFs''), located a couple of degrees away from
M31 (see Table 1) and free
of M31 \ha~emission, were used as comparison spectra. Each direction
was observed for 300 s at a time, alternating between OFFs and ONs.
The data were obtained over a total of five nights during three
different periods separated by almost one year. This was done to take
advantage of the motion of the expected M31 \ha~emission line with respect to
fixed weak atmospheric lines. Table 1 summarizes the details of the
observations.

The WHAM interferometer produced 128$\times$128 pixel ring images from
which spectra were extracted via an annular ring summing
technique; each position within the 1\dg beam is equally sampled by
each spectral element \citep{tuf97}.
The resultant 
spectra were flat-fielded, continuum subtracted, and intensity
calibrated. Signatures of 
cosmic rays and physical defects in the interference filter were
removed with pixel masking.

Figure 2 shows an average of 44 spectra toward all OFF lines of
sight taken on the same night, free of 
M31 emission.  The bright feature in the red side of the spectrum is a
well-known OH atmospheric 
line at 6553.6 \AA, which we used for precise wavelength calibration. 
The feature near -480 \kms~is a Fabry-Perot ghost from the bright geocoronal 
\ha~line near 0 \kms; the other features are extremely weak,
unidentified atmospheric lines within the spectral window. 
The strengths of all of these lines vary both
temporally and spatially on the sky, which make the task of
identifying the weak \ha~emission line from M31 formidable.

To isolate the emission solely due to M31, the spectra obtained in the
OFF directions were subtracted from the ON spectra.
Due to the complex nature of the
intensity variations of the weak atmospheric features, only a subset
of the OFF spectra were subtracted from the ONs.
This subset satisfied the 
criteria that the resultant spectra formed a flat baseline around the region of
expected M31 emission.  This process
produces accurate spectra if the OFFs possess no significant
extra-terrestrial emission features within the
spectrum. The potentially incomplete subtraction of the atmospheric
lines is the largest source of uncertainty in our results.

\section{Results}

Beam 1, positioned at the center of M31, reveals significant
\ha~emission. 
However, due to the dynamical location of the
line of sight, the width of the line extended beyond the
$\pm$100 \kms~range of the spectrum.
Hence, an accurate spectral baseline and estimate of
the emission line strength was not possible.
For the remaining ON beam positions, the expected emission line width
should be relatively narrow, comparable to the thermal and turbulent
motion of the emitting gas.
This is confirmed by observations of M31's flat rotation curve toward
these directions, as seen in the 21 cm spectrum along the major axis
\citep{cra80}.

The spectra taken toward Beam 2 also reveals
significant \ha~emission, as seen in Figure 3, with an intensity of
I$_{H\alpha} = \bti R$.  
The error in this measurement is due to the uncertainty in the spectral
baseline generated by the incomplete subtraction of the weak
atmospheric features.
Toward the line of sight off of the stellar disk, but within the
HI disk (Beam 3), \ha~emission is detected near the velocity
of the HI, with I$_{H\alpha} = \bri R $ (see Figure 4).
Figure 1 shows that this emission arises far from the
optical stellar disk.

Toward the two lines of sight just off the edge of the HI disk (Beams
4 \& 5), no significant \ha~emission was detected.
Upper limits of $I_{H\alpha} \leq \uplimone R$ and $I_{H\alpha}
\leq \uplimtwo R$ were determined for Beams 4 and 5, respectively
(see Figure 4).
These are the strengths of the strongest
emission that can be placed at the HI velocity, with the same
width as the \ha~emission detected in Beam 3, and not be
discernable in the spectra.
Note that the scatter in the spectra toward Beam 5 is considerably
larger than for Beam 4. This increase in the scatter arises from
intrinsically larger variations in the atmospheric lines in this
direction.
A more accurate removal of these lines toward Beam 4 has provided a
stronger upper limit, which we adopt as the largest \ha~emission
strength allowable off the edge of the HI disk in M31.
Although the data indicate a somewhat lower limit, we have concluded
that the variations in the atmospheric lines, which produce incomplete
subtraction of these features in the ON-OFF
spectra, could hide a signal up as large as $\uplimone R$.
We note that the 2$\sigma$ uncertainty due to 
Poisson noise in the data is much lower, at $0.004 R$.

Other deep CCD imaging searches for faint \ha~emission in external galaxies
\citep{zur00,fer98a,mar01,wal92} are 
sensitive to signals ranging from $\approx 0.5$ to $100 R$, with typical 
sensitivities of tens of Rayleighs.
Our ability to detect emission down to $\sim 0.01 R$ is a result of
the unique dual-etalon, high resolution Fabry-Perot design of WHAM,
combined with the power of the ON-OFF technique.

\section{Discussion and Conclusions}

One of the original motivations for these deep \ha~observations of M31
was to search for HII extending beyond the radius of the HI cutoff. No
such extended emission was found down to our detection limit of
$\uplimone R$. 
This corresponds to a face-on emission measure $EM_{\bot} = 0.016$ cm$^{-6}$
pc for a temperature of 10$^4~$K, and a column recombination rate of
$1.2 \times 10^4$ hydrogen recombinations cm$^{-2}$~s$^{-1}$.
To calculate the emission measure and recombination rate we estimated
the extinction of the \ha~photons as they pass through
the Milky Way along each of the lines of sight.
We used the Galactic neutral hydrogen column
density $N_{HI}$ toward each direction \citep{har97}, and
the relationship between $N_{HI}$ and $\tau_{\lambda}$ given by
\citet{boh78} and \citet{mat90} to increase I$_{H\alpha}$ by 
$\approx$ 30\% over the observed I$_{H\alpha}$.
The extinction within the outer disk of M31 is negligible
as compared to the foreground Galaxy \citep{cui01}, and we
therefore have not corrected for that effect.

On the other hand, we did detect ionized hydrogen in the extended HI
disk of M31 (Beam 3). One possible source of this ionization is young,
massive O stars within the HI layer.
\citet{cui01} recently completed a deep, \emph{V} and \emph{I} band
photometric imaging study of a region 
within our Beam 3, and did find evidence of young B stars.
However, they also mention that while an estimate of the threshold
for very massive star formation in this region is difficult to
ascertain, it is unlikely that the gas disk surface density exceeds
the critical density \citep{mar01}.
Therefore, while deep \ha~imaging studies of galaxies of
\emph{later-type} than 
M31 (Sb) have revealed sites of massive star formation at comparably
large galactocentric distances \citep{fer98a,fer98b},
it seems unlikely that the observed \ha~emission toward Beam 3
arises from photoionization from massive O stars. 

Another possible source of this HII is an external flux of Lyman
continuum radiation.  Since the HI fills the WHAM beam for position 3
(see \citet{cra80}),
the observed \ha~intensity of $0.05 R$ provides a direct estimate of
the required ionizing flux \citep{mal93}.
Assuming that the HI absorbs all of the incident Lyman continuum
photons and that the gas is optically thin to \ha, we find that
an ionizing flux incident $\Phi_0 = \phihi \times
10^{4}~$~photons~cm$^{-2}$~s$^{-1}$ on each side of M31 would account
for the \ha.
We derived this value assuming case B and an electron
temperature $T_e = 10^4$ K, making the small extinction correction
described above, and taking into account the 77\dg inclination of
M31's disk from the line of sight. However, since additional sources
of ionization may be present, e.g. B stars, planetary nebulae, or hot
white dwarf stars in the outer disk and halo of M31  \citep{cui01},
this value of $\Phi_0$ must be considered an upper limit on the
external ionizing flux.
Other authors have derived $\Phi_0$
with various techniques combining both observations and theory and
probing a variety of extragalactic environments (see Table 1 of
\citet{shu99}). 
We find that our value of $\Phi_0 = \phihi \times
10^{4}~$~photons~cm$^{-2}$~s$^{-1}$ is among the lowest of these
reported values.

Our upper limit on $\Phi_0$ near M31 implies that a spherical,
isolated HI cloud bathed only in an \emph{isotropic} intergalactic
radiation field of the Local Group 
would have an \ha~intensity $I_{H\alpha} \leq \haiso R$. This is
significantly ($\sim$ factor of 3) below the observed intensities of
high velocity HI 
clouds in the Milky Way's halo \citep{tuf98} and in the
Magellanic Stream \citep{wei96}, implying an additional
source of ionizing radiation in the halo of the Milky Way, such as
Lyman continuum photons from O stars escaping the disk
\citep{blan99,dov00}. 
\ha~observations of high velocity HI
clouds may therefore provide important information about the location
of clouds within the Local Group. Specifically, clouds far from a
galaxy (e.g. \citet{bli99}) should have \ha~intensities
$\leq \haiso R$.

\acknowledgments
GJM, RJR, LMH, and SLT acknowledge support from the National Science
Foundation through grant AST 96-19424.
PRM is supported by the National Science Foundation under grant AST
99-00871.

\begin{deluxetable}{ccccccccc}
  \rotate
  \tablewidth{0pt}
  \tablenum{1}
  \tablecolumns{9}
  \tablecaption{Summary of Observations}
  \tablehead{\colhead{} &
    \colhead{} &
    \colhead{} &
    \colhead{} &
    \multicolumn{2}{c}{Distance} &
    \colhead{I$_{H\alpha}$} &
    \colhead{v$_{H\alpha}$\tablenotemark{d}} &
    \colhead{FWHM} \\
    \colhead{Beam \#} & \colhead{ON/OFF} &
    \colhead{$\alpha$\tablenotemark{a}} &
    \colhead{$\delta$\tablenotemark{a}} &
    \colhead{(kpc)\tablenotemark{b}} &
    \colhead{($R_{25}$)\tablenotemark{c}} &
    \colhead{($R$)} & \colhead{(\kms)} &
    \colhead{(\kms)}}

  \startdata
  1 & ON   & 0\h 40\m 00\s & 41\dg 00' 00'' & ...  & ... &  ...  & ... & ... \\
  2 & ON   & 0\h 37\m 33\s & 40\dg 24' 32'' & 10.3  & 0.47 & $\bti$  & -535 $\pm$ 0.5 & 51 $\pm$ 1 \\
  3 & ON   & 0\h 33\m 16\s & 39\dg 17' 54'' & 29.2  & 1.34 & $\bri$  & -521 $\pm$ 1.5& 36 $\pm$ 5 \\
  4 & ON   & 0\h 31\m 26\s & 38\dg 37' 26'' & 39.5  & 1.82 & $\leq \uplimone$ & 520 & 36 \\
  5 & ON   & 0\h 30\m 00\s & 37\dg 55' 00'' & 49.8  & 2.29 & $\leq \uplimtwo$ & 520 & 36 \\
  6 & OFF  & 0\h 08\m 00\s & 38\dg 15' 00'' & ...  & ... & ...  & ... & ... \\
  7 & OFF  & 1\h 00\m 00\s & 38\dg 00' 00'' & ...  & ... & ...  & ... & ... \\
  8 & OFF  & 0\h 20\m 00\s & 41\dg 00' 00'' & ...  & ... & ...  & ... & ... \\
  9 & OFF  & 0\h 55\m 00\s & 41\dg 30' 00'' & ...  & ... & ...  & ... & ... \\
  \enddata
  \tablenotetext{a}{Coordinates given in [B1950] equinox.}
  \tablenotetext{b}{Galactocentric distance, using a distance of
    784 kpc to M31 \citep{sta98}.}
  \tablenotetext{c}{Radius of the $\mu_B=25$ mag arcsec$^{-2}$
    isophote \citep{dev91}.}
  \tablenotetext{d}{Velocity of the \ha~emission with respect to the
    LSR.}
\end{deluxetable}

\clearpage

\begin{figure}[t]
  \includegraphics*[scale=0.75]{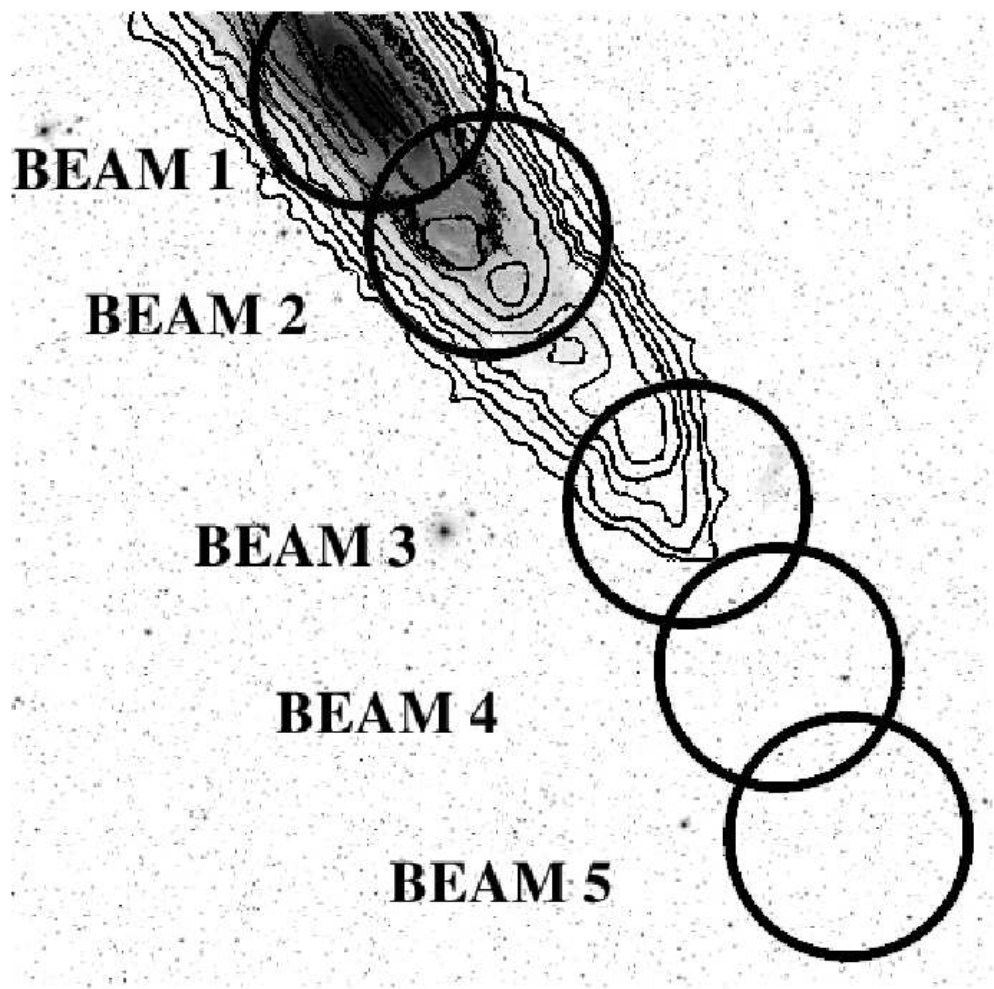}
  \caption{Overlay of an optical image of M31, HI
    contours, and the ON lines of sight considered in this study. The optical
    image is from the DSS-II at STScI. The HI contour diagram is
    adapted from \citet{cra80}. The outermost contour level is at
    N$_{HI}=6.51\times10^{19}$~cm$^{-2}$. Detectable HI emission extends
    beyond this last contour to fill Beam 3.}
  \label{fig:1}
\end{figure}

\begin{figure}[t]
  \includegraphics*[scale=0.75]{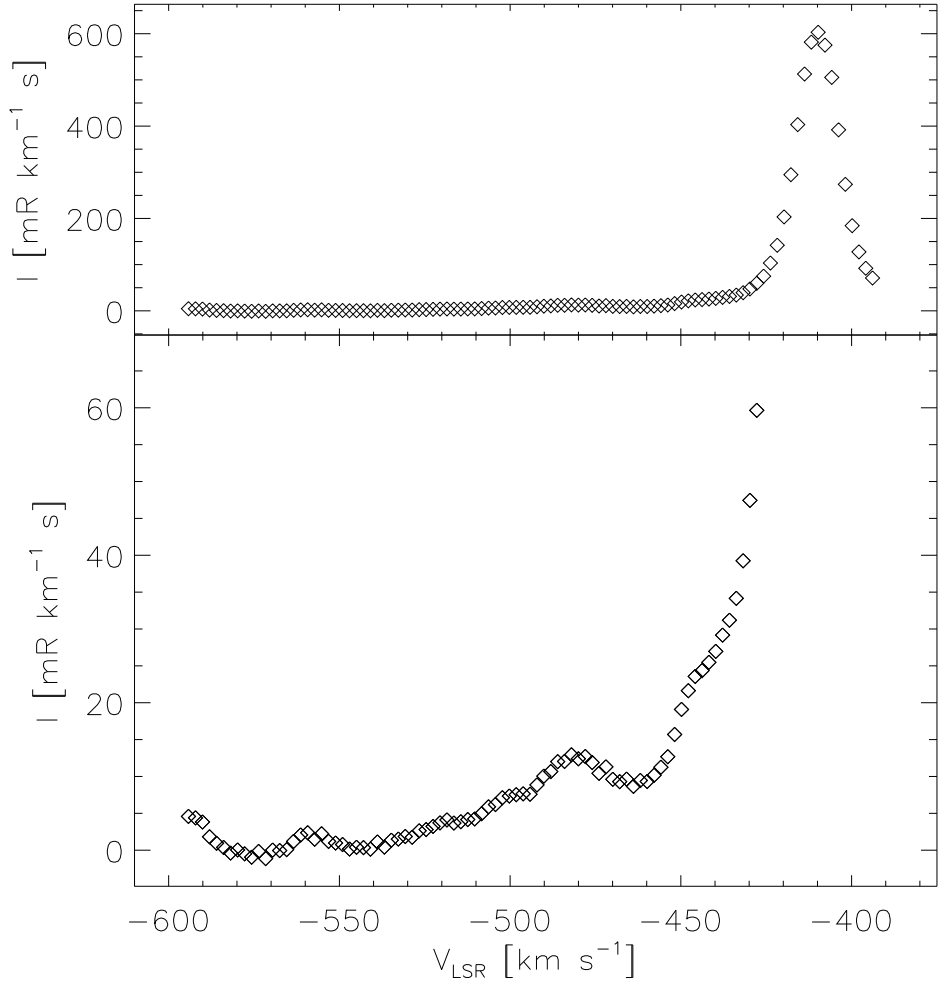}
  \caption{Average spectrum of 44 OFF lines of sight, all taken on
    the same night, for a total integration of 3 h, 40 min. The y-axis
    is given in milli-Rayleighs per \kms. The top graph displays
    the total spectrum; the bottom graph is an expansion of the full
    spectrum at low intensity levels. Note the strong OH line near
    -410 \kms, the geocoronal \ha~ghost at -480 \kms, and several
    weaker atmospheric lines.}
  \label{fig:2}
\end{figure}

\begin{figure}[t]
  \includegraphics*[scale=0.75]{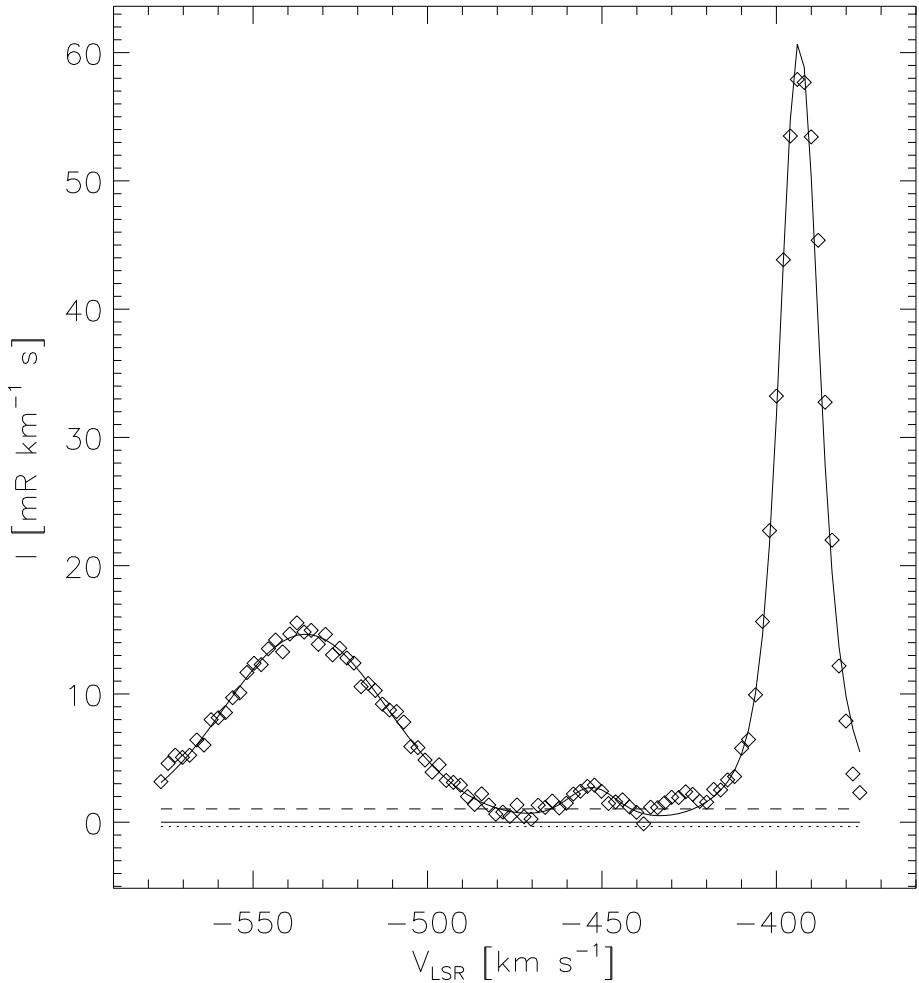}
  \caption{Reduced spectrum of the region toward the southwestern
    stellar disk (Beam 2). The M31 \ha~emission at -535 \kms~has a
    strength of I$_{H\alpha} = \bti R$. The data are represented
    as diamonds and the best fit by a solid line through the
    data. Note the incomplete subtraction of the OH line, and the
    location of the continuum. The straight, solid line is the
    continuum used for the best fit, whereas the dashed and dotted
    lines are upper and lower limits to the continuum, reflected in
    the upper and lower error limits of I$_{H\alpha}$.}
  \label{fig:3}
\end{figure}

\begin{figure}[t]
  \includegraphics*[scale=0.75]{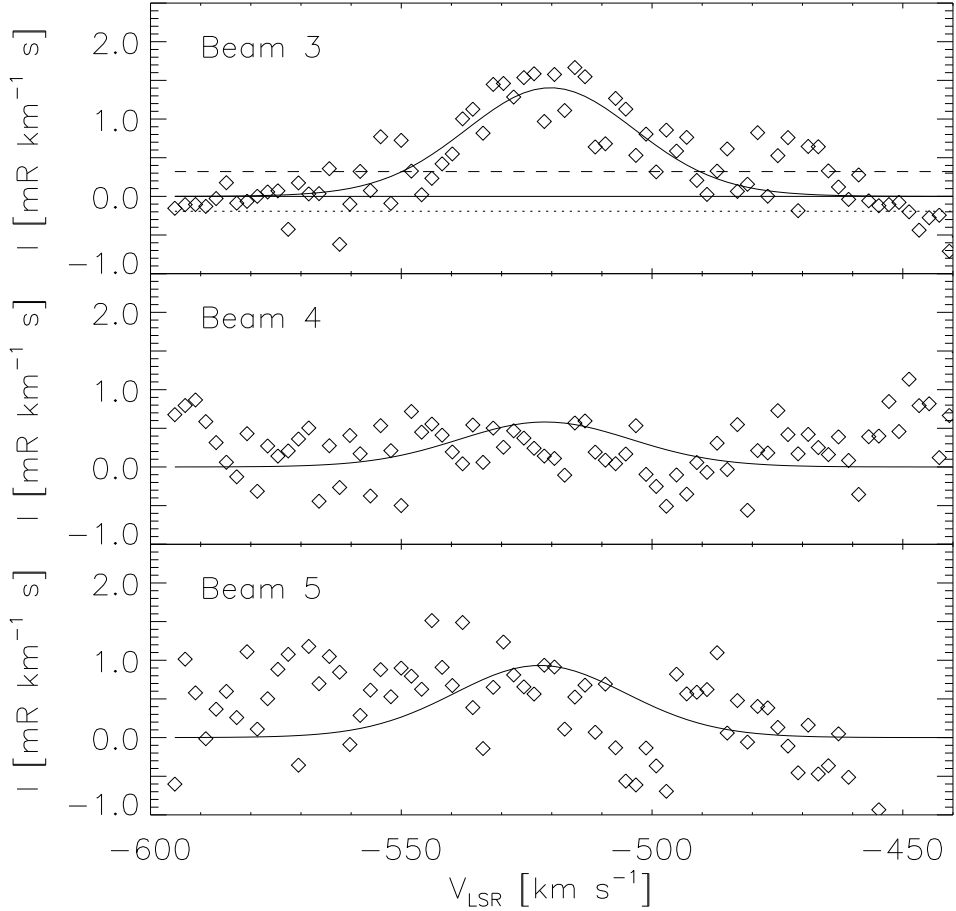}
  \caption{Reduced \ha~spectra away from M31's stellar disk toward Beams
    3, 4 and 5.  \ha~emission from M31 is present in the top plot toward
    Beam 3 at a velocity of -521 \kms, with a strength of I$_{H\alpha}
    = \bri R$ (symbols and lines as in Fig. 3). The middle and
    bottom plots of spectra
    towards Beams 4 \& 5 show no detectable \ha~emission. The solid
    line in these spectra are reasonable lower limits to their continua plus an
    \ha~emission line with the same
    velocity and width as the emission toward Beam 3.
    Lines of strength I$_{H\alpha} = \uplimone R$ and
    I$_{H\alpha} = \uplimtwo R$ are overlayed in Beams 4 and 5, respectively.
    Note the scatter
    in the data and that the vertical scales in all three plots are
    identical. }
  \label{fig:4}
\end{figure}

\end{document}